\documentclass[12pt]{article}
\usepackage{epsfig}
\usepackage{amsmath}
\usepackage{hhline}
\usepackage{amssymb}
\usepackage{times}

\newlength{\dinwidth}
\newlength{\dinmargin}
\newcommand{\GeV}{{\rm\,GeV}}
\newcommand{\gev}{{\rm\,GeV}}
\newcommand{\mev}{{\rm\,MeV}}

\newcommand{\pbinv}{{\rm\,pb^{-1}}}

\newcommand{\pt}{p_{_{\rm T}}}

\newcommand{\GeVSq}{\rm\,GeV^2}

\newcommand{\pqqc}{$\Xi^{--}_{5q}\,$}
\newcommand{\pqqn}{$\Xi^{0}_{5q}\,$}
\newcommand{\ximm}{$X^{--}\,$}
\newcommand{\xipm}{$X^{0}\,$}
\newcommand{\xizero}{$\Xi(1530)^{0}\,$}

\begin{document}

\begin{titlepage}

\begin{flushleft}

DESY 07-045 \hfill ISSN 0418-9833 \\
April 2007
\end{flushleft}

\vspace{2cm}
\noindent

\begin{center}
\begin{Large}

{\bf Search for Baryonic Resonances Decaying to $ {\Xi\pi}$ \\
   in Deep-Inelastic Scattering  at HERA }

\vspace{2cm}

H1 Collaboration

\end{Large}
\end{center}

\vspace{2cm}

\begin{abstract}

\noindent
A search for narrow baryonic resonances decaying into  
$\Xi^{-}\pi^{-}$ or\, $\Xi^{-}\pi^{+}$ and their antiparticles 
is carried out
with the H1 detector using deep inelastic scattering events at HERA
in the range of negative photon four-momentum transfer squared
$2 < Q^2 < 100$ GeV$^2$.
No signal is observed for a new baryonic state in 
the mass range $1600 - 2300 \mev$
in either  the doubly charged or the  neutral 
decay channels.
The known baryon \xizero is observed through its
decay  mode into $\Xi^{-}\pi^{+}$.
Upper limits are given on the ratio of the production rates
of new baryonic states, such as the hypothetical pentaquark states
\pqqc or \pqqn, relative to the \xizero baryon state.

\end{abstract}

\vspace{1.5cm}

\begin{center}
To be submitted to Phys. Lett. {\bf B}
\end{center}

\end{titlepage}

%
\begin{flushleft}

A.~Aktas$^{10}$,               
C.~Alexa$^{10,49}$,            
V.~Andreev$^{24}$,             
T.~Anthonis$^{4}$,             
B.~Antunovic$^{25}$,           
S.~Aplin$^{10}$,               
A.~Asmone$^{32}$,              
A.~Astvatsatourov$^{4}$,       
S.~Backovic$^{29}$,            
A.~Baghdasaryan$^{37}$
% ...

\bigskip{\it
% ...
$ ^{2}$ Vinca  Institute of Nuclear Sciences, Belgrade, Serbia \\
% ...
$ ^{4}$ Inter-University Institute for High Energies ULB-VUB, Brussels;
          Universiteit Antwerpen, Antwerpen; Belgium$^{ c}$ \\
% ...
$ ^{10}$ DESY, Hamburg, Germany \\
% ...
$ ^{24}$ Lebedev Physical Institute, Moscow, Russia$^{ e}$ \\
 $ ^{25}$ Max-Planck-Institut f\"ur Physik, M\"unchen, Germany \\
% ...
$ ^{29}$ Faculty of Science, University of Montenegro,
          Podgorica, Montenegro$^{ e}$ \\
% ...
$ ^{32}$ Dipartimento di Fisica Universit\`a di Roma Tre
          and INFN Roma~3, Roma, Italy \\
% ...
$ ^{37}$ Yerevan Physics Institute, Yerevan, Armenia \\
% ...
$ ^{49}$ Also at National Institute for Physics and Nuclear Engineering,
          Magurele, Bucharest, Romania \\

\bigskip
% ...
$ ^c$ Supported by FNRS-FWO-Vlaanderen, IISN-IIKW and IWT
      and  by Interuniversity
Attraction Poles Programme,
      Belgian Science Policy \\
% ...
$ ^e$ Supported by the Deutsche Forschungsgemeinschaft \\
% ...
}
\end{flushleft}
%


\newpage

\section{Introduction}

A search for narrow baryonic states in the mass range $1600 - 2300 \mev$, 
decaying into the 
doubly charged $\Xi^{-}\pi^{-}$ and
neutral $\Xi^{-}\pi^{+}$ 
final states and their antiparticles\footnote{Unless explicitly mentioned, 
the charge conjugate states are hereafter always implicitly included.}, is presented. 
The search is carried out in deep inelastic $ep$ scattering 
(DIS) at HERA.

Various theoretical approaches~\cite{theory1,theory2,theory3} 
based on Quantum Chromodynamics predict the
existence of exotic baryonic states composed of four valence quarks
and an anti-quark, commonly known as ``pentaquarks''. 
Such states are expected to form a flavour anti-decuplet and are not
explicitly forbidden within the Standard Model.

Several experiments have reported evidence for a narrow resonance 
with a mass around 
$1540\, \mev$ decaying into  $nK^{+}$ and $pK^{0}_{S}$ 
final states~\cite{Hicks:2005gp}.
Such a state could be interpreted as an exotic strange pentaquark
with a minimal quark content of $uudd\bar{s}$, lying in the apex of the
spin $1/2$ (or $3/2$) anti-decuplet.
%
On the other hand, a number of other experiments~\cite{Hicks:2005gp},
including the H1 experiment~\cite{Aktas:2006ic}, 
have reported non-observation of the same state.

Searches for other members of this anti-decuplet are of interest, in particular
for the doubly strange  \pqqc $(ddss\bar{u})$  
and $\Xi^+_{5q}$ $(uuss\bar{d})$ states, which are also 
manifestly exotic.
The NA49  collaboration~\cite{na49}
reported the observation of two baryonic resonances
in fixed targed proton-proton 
collisions at the SPS at the centre of mass energy $\sqrt{s} = 17.2$ GeV,
with masses of $1862\pm 2\mev$ and $1864\pm 5\mev$, and
with widths below the mass resolution of $18\mev$. 
These states can be interpreted as the 
 \pqqc ($S$\,=\,$-2$, $I_3=-3/2$) and
the \pqqn ($S$\,=\,$-2$, $I_3=+1/2$)
members of the isospin $3/2$ quartet $\Xi_{3/2}$ in the anti-decuplet.
These findings have not been confirmed by several other experiments  
\cite{ximinus-neg,Schael:2004nm,Bai:2004gk,Abt:2004tz,Airapetian:2004mi,Chekanov:2005at}.

The search presented here is performed using data 
taken with the H1 detector at HERA.
The $\Xi^{-}$ particles
are identified through
their decay into $\Lambda\pi^{-}$. 
The established baryon \xizero~\cite{pdg06} 
is observed through its decay mode
$\Xi(1530)^{0}\rightarrow\Xi^{-}\pi^{+}$.

%
\section{Experimental Procedure}
\label{method}
\subsection{H1 Apparatus}
\label{detector}
A detailed description of the H1 detector can be found in \cite{h1det}. 
In the following only those detector components important 
for the present analysis are described.

H1 uses a right handed cartesian
coordinate system with the origin at the nominal $ep$
interaction point.
The proton beam direction defines the $z$ axis.
The polar angle $\theta$ is measured with respect
to this axis and 
the pseudorapidity $\eta$ 
is given by $\eta=-\ln {\tan {\frac{\theta}{2}}}$.

The tracks from charged particles are reconstructed in the 
central tracker, whose main components are two cylindrical drift chambers,
the inner and outer central jet chambers (CJCs) \cite{Burger:eb}, complemented by 
a central silicon vertex detector (CST) \cite{Pitzl:2000wz}.
The CJCs are mounted concentrically around the 
beam-line, 
covering the range of pseudorapidity
\mbox{$-1.75 < \eta<1.75$} for tracks coming from the 
nominal interaction point.
The CJCs lie within a homogeneous magnetic field of 
$1.16 \, {\rm T}$ which allows the transverse momentum $\pt$ 
of charged particles to be measured.
Two additional drift chambers (CIZ, COZ) complement the CJCs by precisely
measuring the $z$ coordinates of track segments and hence improve the
determination of the polar angle. 
Two cylindrical multi-wire proportional chambers facilitate
triggering on tracks.
The transverse momentum resolution of the central tracker is
$\sigma(\pt) / \pt \simeq 0.005 \, \pt \, /\,\GeV \, \oplus 0.015$~\cite{Kleinwort}.
Charge misidentification is negligible
for particles originating from the primary vertex and having 
transverse momenta in the range relevant to this analysis. 

The tracking detectors are surrounded by a Liquid Argon calorimeter (LAr) in 
the forward and central region ($-1.5 < \eta < 3.4$) and by a lead-scintillating 
fibre calorimeter (SpaCal) in the 
backward region~\cite{Appuhn:1996na} ($-4 < \eta < -1.4$)\@. 
The SpaCal is optimised for the detection of the scattered 
electron\footnote{Herein, the term ``electron'' is used generically to 
refer to both electrons and positrons.} 
in the DIS kinematic range considered here.
A planar drift chamber, positioned in front of the SpaCal, 
improves the measurement of the electron polar angle and
is used to reject neutral particle background from photons.
The kinematics of the hadronic final state are reconstructed 
using an algorithm which combines information from the
central tracker, the SpaCal and the LAr calorimeter. 
The DIS events studied in this paper are triggered 
by an energy deposition in the SpaCal,
complemented by signals in the CJCs and the multi-wire proportional 
chambers in the central tracker. 

The luminosity is determined from the rate of Bethe-Heitler processes 
$ep \rightarrow ep\gamma$, measured using a calorimeter located
close to the beam pipe at $z=-103~{\rm m}$ in the backward direction.

%
\subsection{Selection of DIS Events}
The analysis is carried out using data corresponding to an integrated 
luminosity of ${\cal L}=101 \pbinv$, taken in the years
$1996/1997$ and $1999/2000$. During this time HERA collided 
electrons or positrons at an energy of $27.6\gev$ with protons at 
$820\gev$ ($1996 / 1997, 24.8$\,${\rm pb^{-1}}$) 
and $920\gev$ ($1999 / 2000, 75.7$\,${\rm pb^{-1}}$).

The event kinematics are reconstructed from the 
energy  and the  polar angle of the scattered electron, which
is required to be identified in the SpaCal
with an energy above $8\gev$.
The negative four momentum transfer squared $Q^2$ of the exchanged virtual 
photon and the \mbox{inelasticity $y$} 
are required to lie in the ranges 
$2 < Q^2 < 100\GeVSq$ and
$0.05 < y < 0.7$.
The difference between the total energy $E$ and the longitudinal 
component of the total momentum $p_z$, calculated 
from the electron and the hadronic final state, is restricted
to $35 < E - p_{\rm z} <70\gev$. This requirement suppresses  
photoproduction background, in which the electron escapes detection and 
a hadron fakes the electron signature.
Events are accepted if the
$z$ coordinate of the event vertex, reconstructed using the 
central tracker, lies within $\pm 35{\rm\,cm}$ of
the mean position for $ep$ interactions. 

%
\subsection{Selection of the Baryon Candidates}
\label{lambda}
The hypothetical doubly charged $X^{--}$ and the neutral  $X^{0}$ baryon states
are identified by complete
reconstruction of their respective decay chains through $\Xi^-$ and
$\Lambda$ baryons into pions and protons, according to 

\begin{eqnarray}
X^{--} & \to & \Xi^- \pi^- \to [\Lambda \pi^-] \pi^-  \to [(p \pi^-)\pi^- ] \pi^- \\
X^{0} & \to & \Xi^- \pi^+ \to [\Lambda \pi^-] \pi^+  \to [(p \pi^-)\pi^- ] \pi^+ \,.
\label{eq:decay}
\end{eqnarray}
The decay daughter particles are fitted
in three dimensions to their respective decay vertices \cite{lg-fit}, 
which are referred to 
as the tertiary ($\Lambda$-decay) and the secondary ($\Xi$-decay) vertex.
The analysis is based on charged particles measured in the central 
region of the H1 detector with a minimal transverse momentum $\pt$ of $0.12\gev$.
These (non vertex-fitted) tracks may in addition be kinematically 
constrained to the primary vertex, in which case they are specifically 
referred to as primary vertex-fitted tracks.
The baryon selection is chosen to optimise the signal for the standard
baryon \xizero.

In the first step, the $\Lambda$ baryons  are identified by their
charged decay mode,  $\Lambda\to p\pi^{-}$, using pairs of oppositely charged
tracks.
The track with the higher momentum is assigned the proton mass.
The particles are fitted to the tertiary vertex
and the $\Lambda$ candidates are retained if the
vertex fit probability is above $1$\%.
To reduce background, the two-dimensional separation (in the 
plane transverse to the beam line)
of the tertiary and the primary vertex is required to be larger than 
\mbox{$7.5{\rm\,mm}$}.  
The transverse momentum of the $\Lambda$ 
is required to be larger than $0.3\gev$.
The contamination from $K^0_S \to \pi^{+}\pi^{-}$ decays
is suppressed by excluding candidates with a 
$\pi \pi$ invariant mass in a $\pm 10\mev$ ($\sim 2\sigma$) window
around the nominal $K^0_S$ mass. $\Lambda$ candidates are retained 
for the further steps if their invariant $p \pi$ mass is within 
$\pm 8\mev$ of the nominal $\Lambda$ mass, 
which is subsequently
assigned to the selected $\Lambda$ candidates.

\begin{figure}
\begin{center}
\includegraphics[width=79mm]{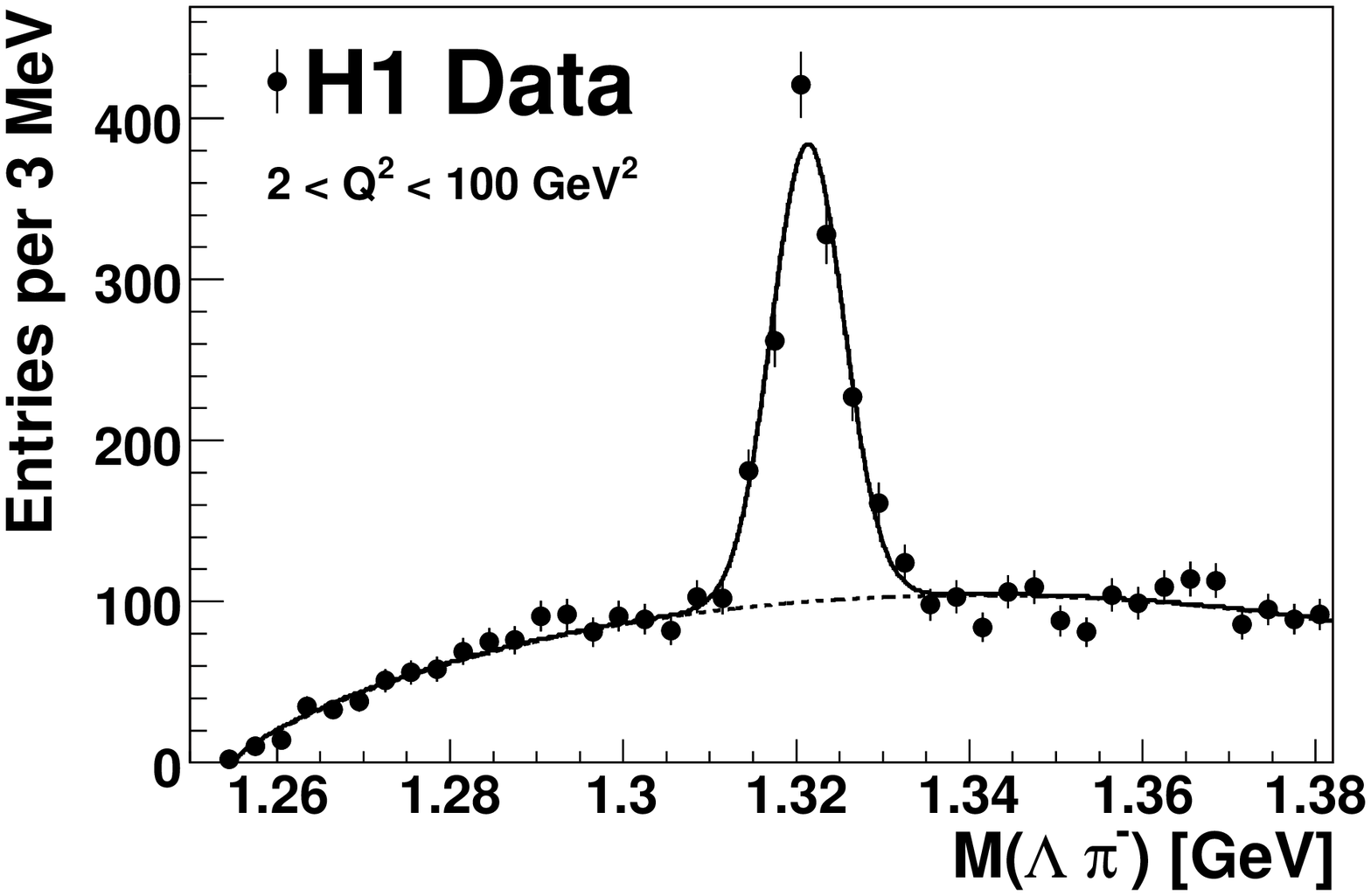}
\includegraphics[width=79mm]{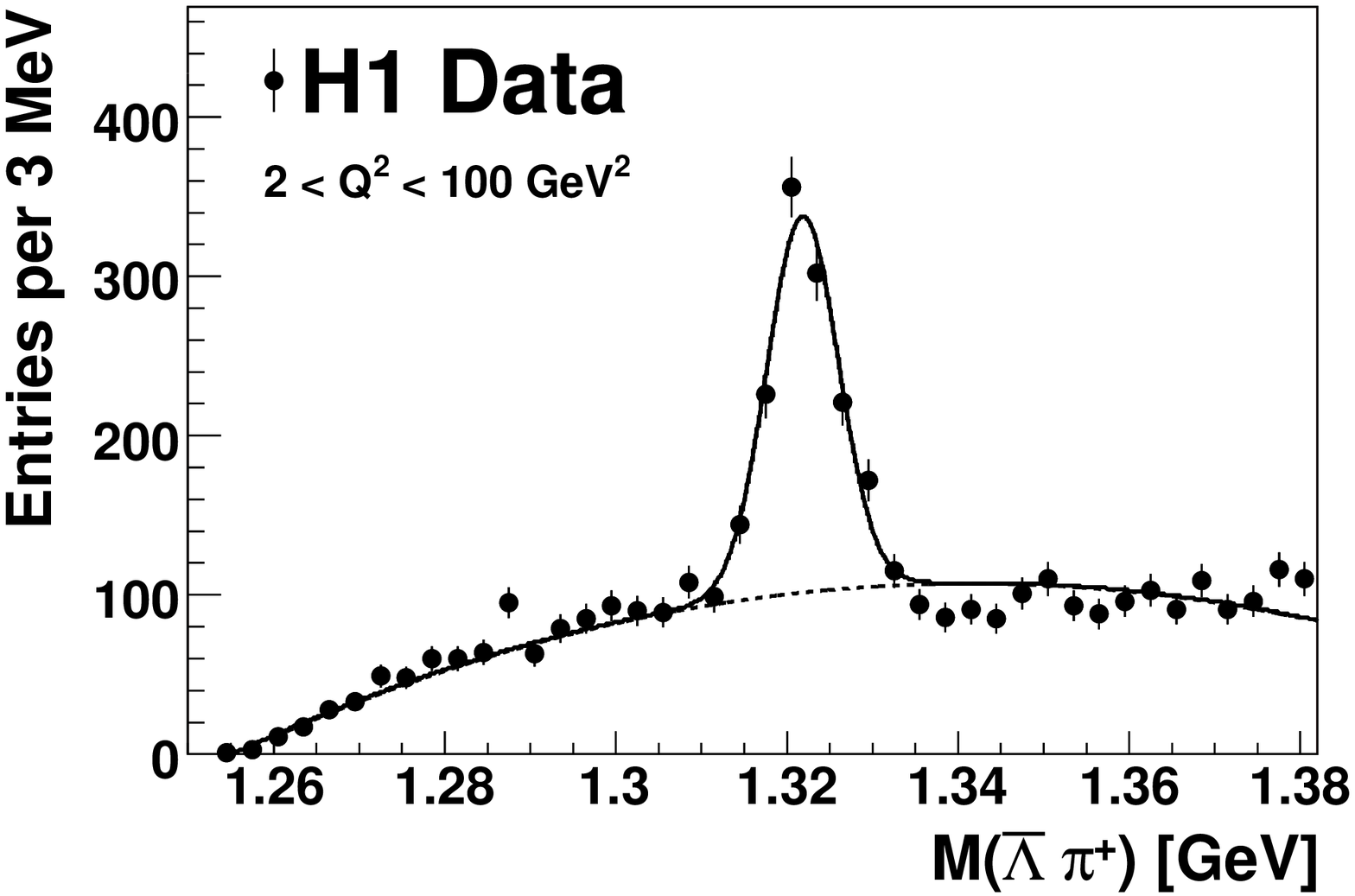}
\setlength{\unitlength}{\textwidth}
\begin{picture}(0,0)
   \put(-0.07,0.31){\bfseries a)}
   \put(0.43,0.31){\bfseries b)}
\end{picture}
\caption{The invariant mass spectra for 
a) $\Lambda\pi^{-}$ and b) $\overline{\Lambda}\pi^{+}$ 
particle combinations.
The solid lines show the result of a fit to the data
using a Gaussian function for the $\Xi^-$ signal and a background 
function as defined in equation~\ref{bgeqn} 
(with $m_{\Xi}$ replaced accordingly by $m_{\Lambda}$),
while the dashed lines indicate the background function only.
} 
\label{xi_mass_spec}
\end{center}
\end{figure}

In the second step,\ $\Xi^{-}$ candidates are formed by
combining each of the $\Lambda$ candidates with a negatively charged 
track assumed to be a pion.
It is required that: 
the $\Lambda$ candidates and the negatively charged tracks be fitted to a secondary
vertex with a probability above $0.1$\,\%; 
the two-dimensional distance of closest approach in the transverse plane ($DCA$) 
of the $\Xi^-$ candidates with respect to the primary vertex
be smaller than \mbox{$2.5{\rm\,mm}$};
the angle between the secondary and the tertiary vertex 
vectors\footnote{A vertex vector is defined as the vector pointing from 
the primary vertex to the given vertex.}
be less than \mbox{$0.6{\rm\,rad}$}.
The invariant mass spectra $M(\Lambda \pi^{-}$) and $M(\bar{\Lambda}\pi^{+}$) 
of all candidates passing these criteria are shown in figure~\ref{xi_mass_spec}. 
Clear signal peaks are observed around the 
nominal $\Xi^{-}$ mass. $\Xi^{-}$ candidates are retained for the third step if their 
invariant mass is within $\pm 15\mev$ of the nominal $\Xi^{-}$ mass,
which is subsequently
assigned to the selected $\Xi^{-}$ candidates.

In the third step, $X^{--/0}$ baryon candidates
are formed by combining each $\Xi^{-}$ candidate
with one additional primary vertex-fitted track assumed to be a pion. 
For these combinations it is required that:
the additional $\pi^{\pm}$ track have
a significance ($DCA/\sigma_{DCA}$) to the primary 
vertex of less than $4.0$;
the transverse momentum of the $X^{--/0}$ candidates
be larger than $1.0\gev$ and their pseudorapitity be within $|\eta| < 1.5$.

%
\subsection{Simulation of Baryonic States}
\label{models}
To estimate the acceptance, the efficiency and the resolution for the detection of a 
hypothetical baryon state,
a Monte Carlo simulation based on the PYTHIA \cite{Sjostrand:2000wi}
event generator is used, incorporating 
the Lund string model of fragmentation \cite{lund}.
The kinematic distributions of strange $S$\,=\,$-1$ hadrons in DIS data have been 
found to be reasonably well described \cite{strangenessdis} by the PYTHIA
simulation.
A generic \ximm baryon state is introduced in the simulation by changing the mass 
of the known
$\bar{\Delta}^{--}$ antibaryon to values in the required range
 from $1600 - 2300 \mev$ and
having it decay into $\Xi^- \pi^-$ final states. 
Because most theoretical predictions
and the measurements~\cite{Hicks:2005gp}  
suggest a very small intrinsic width $\sigma \lesssim 1\, \mev$
for pentaquark states, 
the width for the \ximm is assumed to be zero in the simulation.
Due to lack of knowledge of the pentaquark production mechanism, it 
is assumed that the production kinematics of the \ximm 
are similar to those of the standard \xizero baryon. 
The generic \xipm state is introduced by changing the mass of the known \xizero baryon 
to values in the required range from $1600 - 2300 \mev$,
and letting it decay into $\Xi^- \pi^+$.
%
All the generated events are passed through a detailed simulation of the 
H1 detector response based on the GEANT program~\cite{geant} 
and the same reconstruction and analysis algorithms as used for the data.


\section{Mass Spectra and Limit Calculation}

\begin{figure}
\begin{center}
\includegraphics[width=79mm]{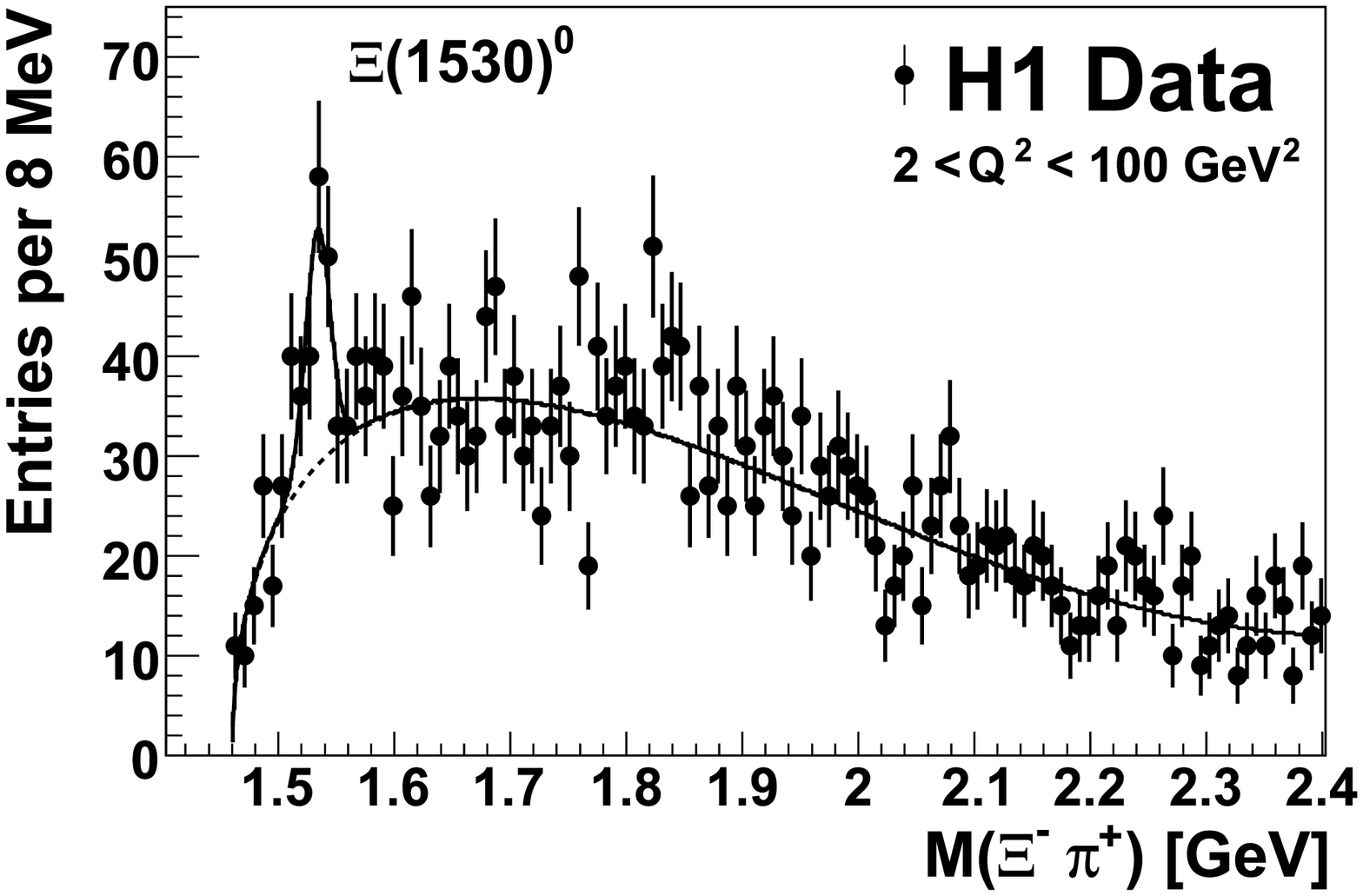}
\includegraphics[width=79mm]{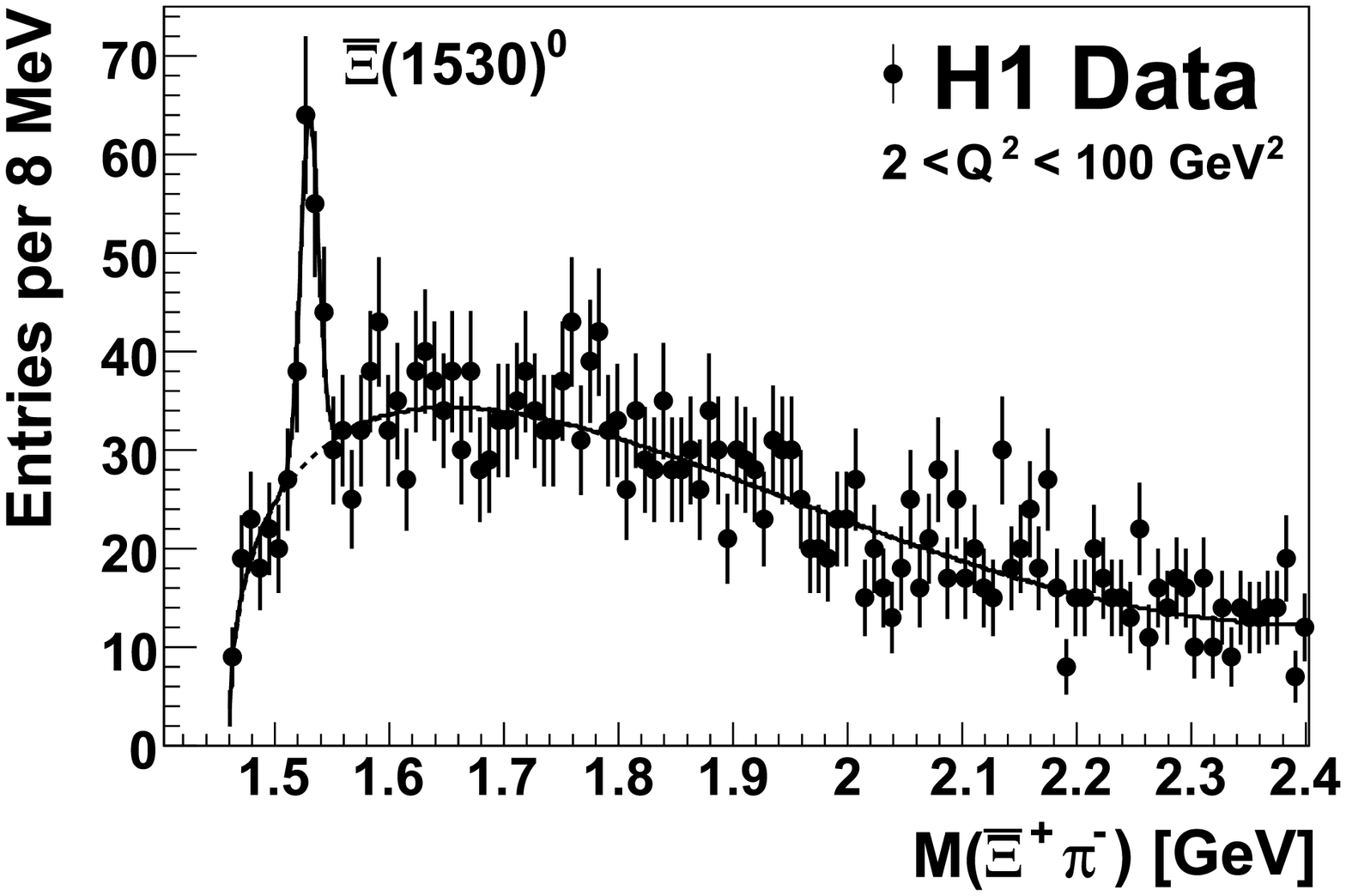} \\
\includegraphics[width=79mm]{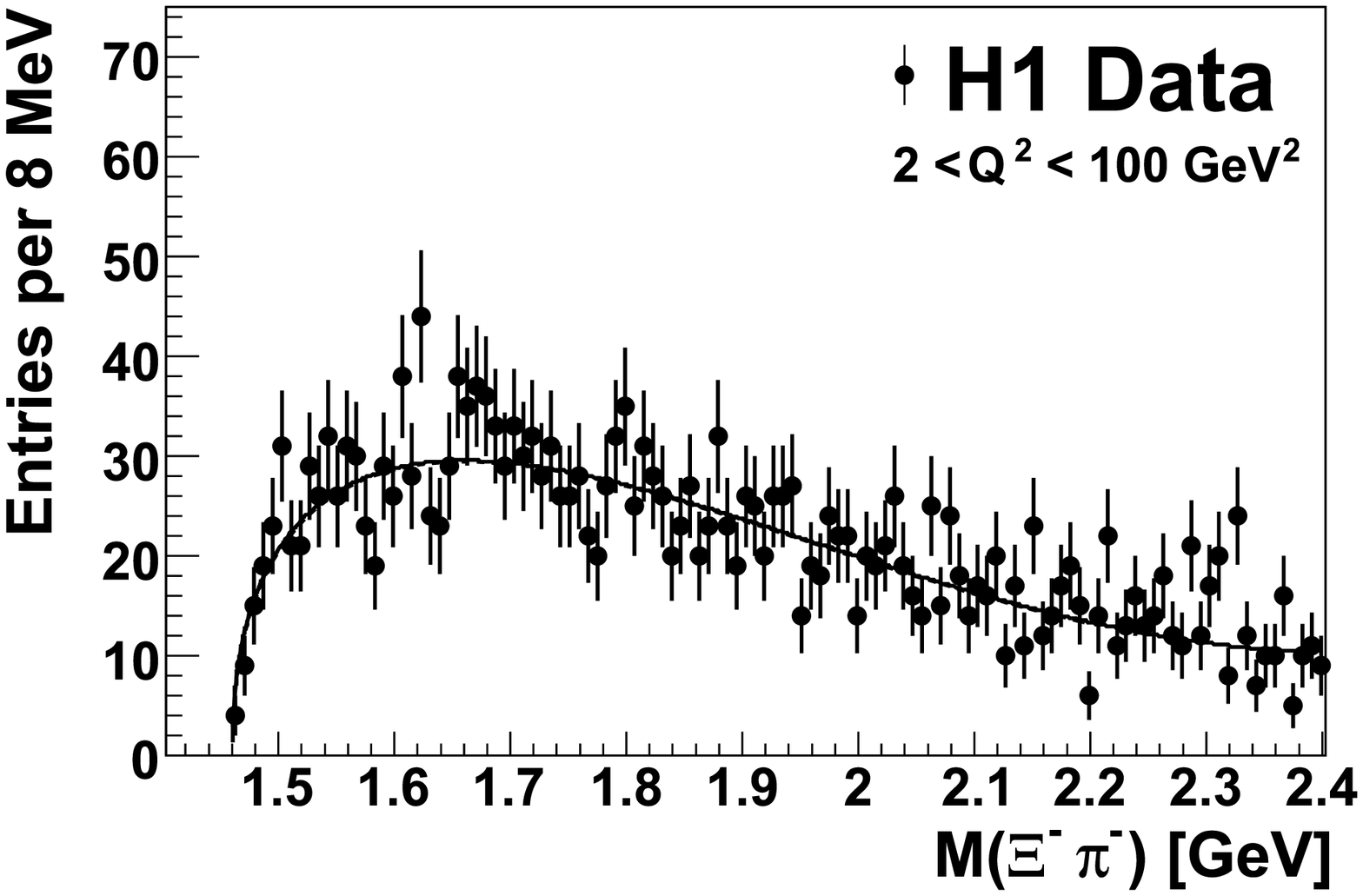}
\includegraphics[width=79mm]{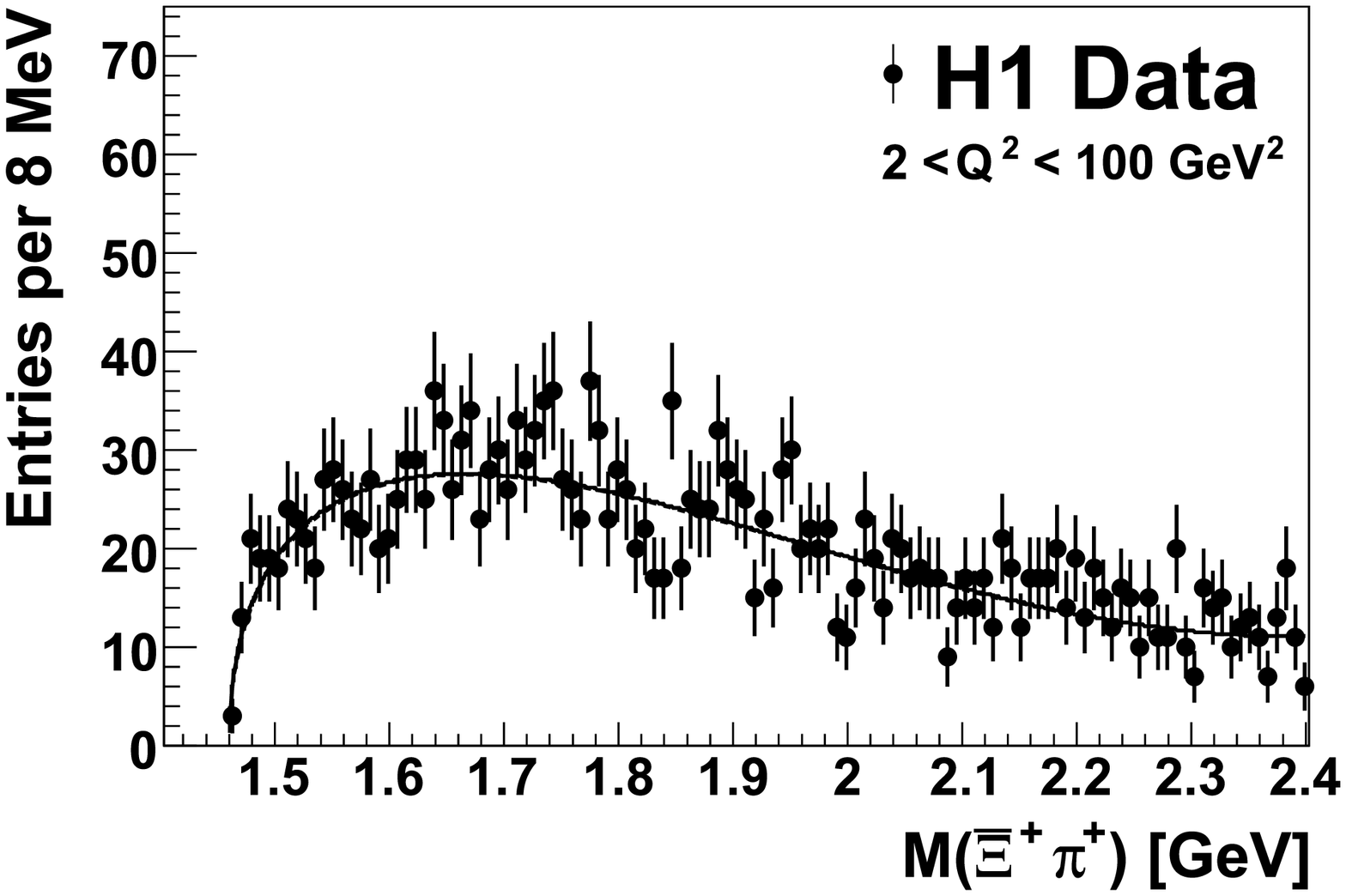} \\
\begin{picture}(0,0)
   \put(-66.5,105){\bfseries a)}
   \put(13.,105){\bfseries b)}
   \put(-66.5,50){\bfseries c)}
   \put(13.,50){\bfseries d)}
\end{picture}
\caption{The  invariant mass spectra of
the $\Xi \pi$ particle combinations for the charge combinations
a) $\Xi^- \pi^+$, b) $\bar\Xi^+ \pi^-$, c) $\Xi^- \pi^-$ and d) $\bar\Xi^+ \pi^+$.
For the purpose of illustration, the solid lines in a) and b) show the result of a fit to the data 
using a Gaussian function for the \xizero signal and a background 
function (dashed lines) as defined in equation~\ref{bgeqn}.
The same background parametrisation is used in c) and d).}
\label{penta_mass_spec}
\end{center}
\end{figure}

The resulting invariant mass spectra for each of the
two neutral ($\Xi^{-}\pi^{+}$, ${\bar\Xi}^{+}\pi^{-}$)
and the two doubly charged combinations 
($\Xi^{-}\pi^{-}$, ${\bar\Xi}^{+}\pi^{+}$)
are shown separately in figure~\ref{penta_mass_spec}.
In both neutral spectra
the signal of the well known \xizero state is observed. 
The sum of the two neutral and of the two doubly charged 
mass spectra are shown in the upper part of figures 
3 and 4, respectively.

\begin{figure}
\begin{center}
\includegraphics[width=105mm]{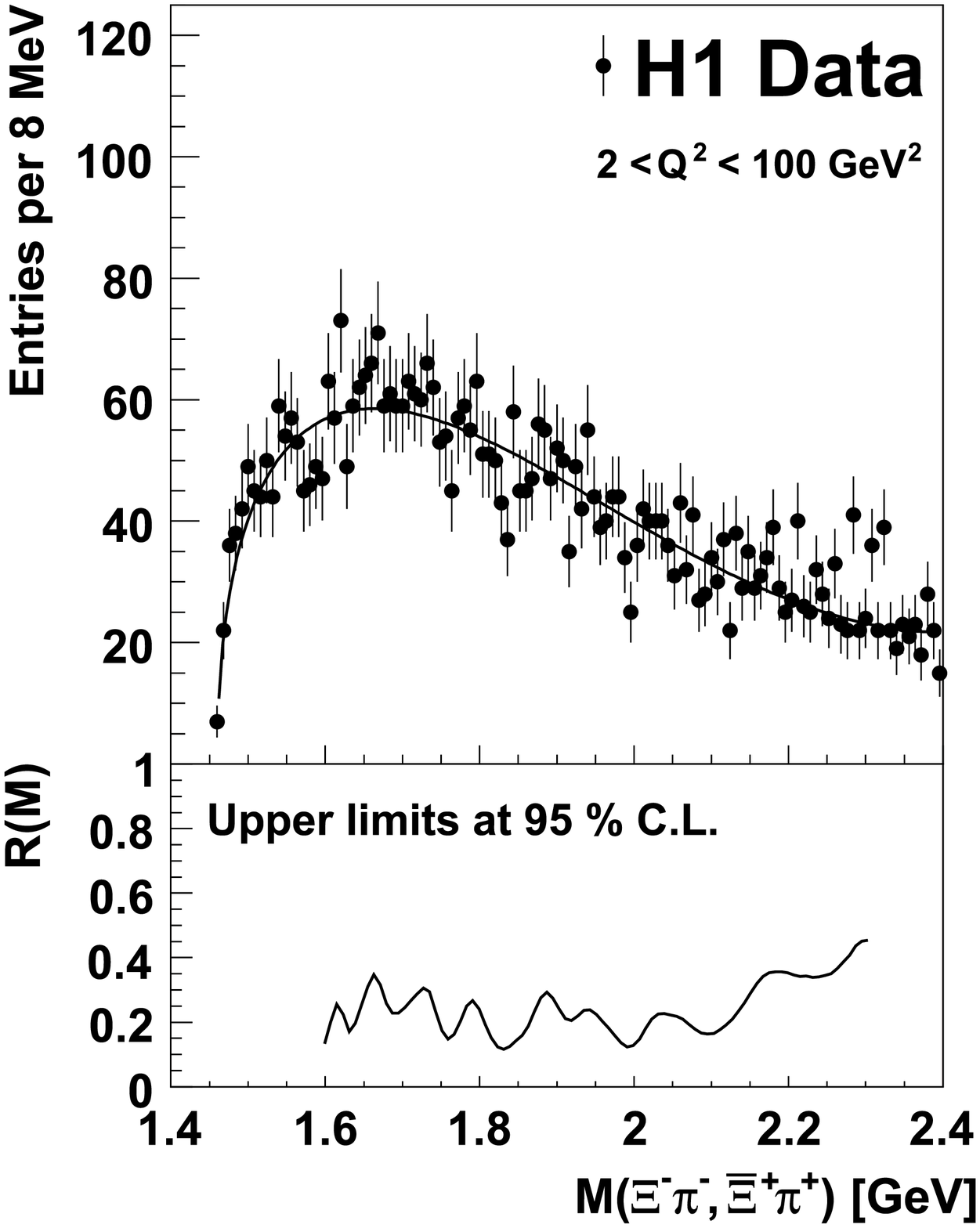}%
\begin{picture}(0,0)
\end{picture}
\caption{The  invariant mass spectrum for
the doubly charged combinations $\Xi^- \pi^-$ and $\bar\Xi^+ \pi^+$ (upper part).
The solid line shows the result of a fit to the data using  
a background function as 
defined in equation~\ref{bgeqn}.
The lower part shows the $95\%$ C.L. upper limit on the ratio $R(M)$ 
as a function of the mass $M$, as
defined in equation~\ref{eq:ratio}.
}
\label{pentamm_limits}
\end{center}
\end{figure}

\begin{figure}
\begin{center}
\includegraphics[width=105mm]{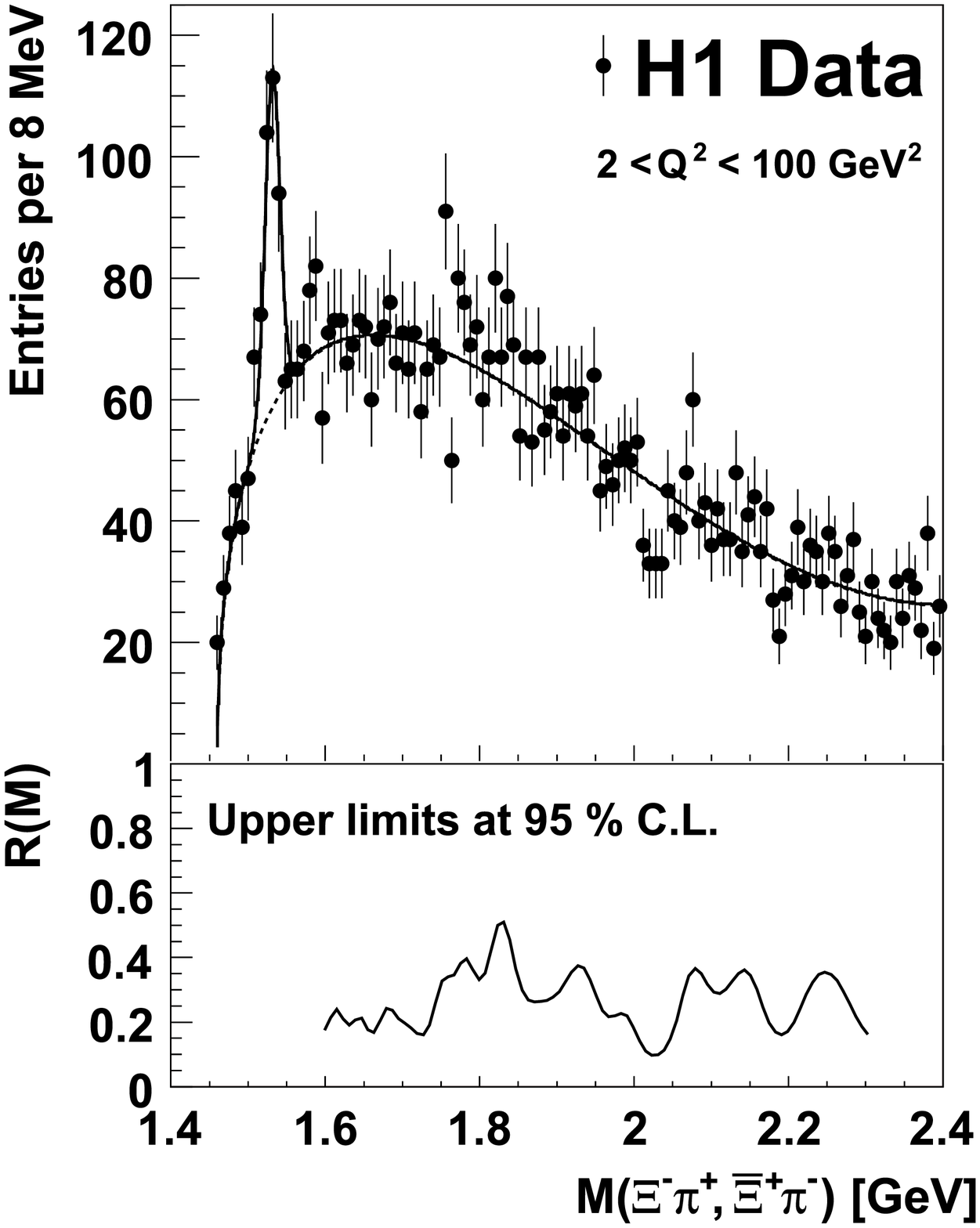}%
\begin{picture}(0,0)
\end{picture}
\caption{The invariant mass spectrum for
the neutral combinations $\Xi^- \pi^+$ and $\bar\Xi^+ \pi^-$ (upper part). 
The solid line shows the result of a fit to the data 
using a Gaussian function for the \xizero signal and a background 
function (dashed line) as defined in equation~\ref{bgeqn}.
%
%
The lower part shows the $95\%$ C.L. upper limit on the ratio $R(M)$ 
as a function of the mass $M$, as
defined in equation~\ref{eq:ratio}.
}
\label{pentamp_limits}
\end{center}
\end{figure}

The ratio of the invariant mass spectra of the
neutral to the charged states
is consistent with being flat in the mass region above the 
\xizero mass. 
This justifies the use of the same background shape function 
for both the doubly charged and the neutral combinations.
%
A simultaneous fit of the neutral and the doubly charged mass spectra, 
shown in figures~\ref{pentamm_limits} and \ref{pentamp_limits}, 
is performed using a function $F$, that contains a
Gaussian function $G$ 
for the signal \xizero baryon and 
a function $B$ for the background shape,
according to

\begin{equation}
\label{bgeqn}
F = G + (1 + P_0)B; \quad  
B(M)=P_{1}(M-m_{\Xi}-m_{\pi})^{P_{2}} \times (1 + P_{3}M + P_{4} M^2).
\end{equation}
\noindent
Here, $M$ denotes the $\Xi\pi$ invariant mass and $m_{\Xi}$ and 
$m_{\pi}$ the masses
of the $\Xi$ and the $\pi$, respectively. The normalisation, the 
central value and the width of the Gaussian function $G$ 
and the parameters $P_{i}$ are left free in the fit. 
$P_0$ represents the relative normalisation of the neutral to the
doubly charged combinations and is set to zero for the neutral combinations.
The fit yields a total 
of $163\,\pm\,24\,(stat.)$ \xizero baryons. 
The reconstructed mass of $1532\,\pm\,2\,(stat.)\mev$
is consistent with the nominal value \cite{pdg06}.
The measured width of $9.4\,\pm\,1.5(stat.)\mev$ is 
in agreement with the detector simulation.

No signal of a new baryonic state
is observed above the \xizero mass 
in either the neutral or  the doubly charged 
mass spectra  (figures~\ref{pentamm_limits} and \ref{pentamp_limits}).
The resonance search can also be performed
relative to the observed signal of the
known \xizero baryon, using the
ratio $R$, which is defined as

\begin{equation}
\label{eq:ratio}
R(M) = \frac{N^{res}(M,q)} {N(1530,0)}
\times \frac{\epsilon(1530,0)}{\epsilon(M,q)}.
\end{equation}
\noindent
$N(1530,0)$ represents 
the number of observed $\Xi(1530)^{0} \to {\Xi}^{-}\pi^{+}$  and
their antiparticle decays.
$N^{res}(M,q)$ describes the estimated number of resonance decays 
depending on the mass $M$ and the charge $q$ of the final state,
which is derived from the difference between the observed spectra
and the expected background contribution.
The background distribution is taken to be the fitted 
function given by equation~\ref{bgeqn}.
For the calculation of $N^{res}$, the 
mass distribution of the signal is assumed to be  
a Gaussian function with a mean $M$ and a 
mass-dependent width $\sigma(M)$ 
corresponding to the experimental mass resolution.
This width $\sigma(M)$ varies from $6.8$ to $22.8\mev$ in the mass range 
considered here, as obtained from the Monte Carlo simulation. 
The term $\epsilon(M,q)$ describes
the reconstruction efficiency of the $\Xi\,\pi$ final state 
as determined by Monte Carlo simulation, 
and it depends 
on the $\Xi\pi$ invariant mass $M$ and on the charge $q$ of the final state.
Accordingly, $\epsilon(1530,0)$ represents the reconstruction efficiency for
the neutral \xizero\ baryon. 
The ratio of efficiencies
 in equation~\ref{eq:ratio} compensates for the small difference in the 
reconstruction efficiencies of the \xizero\ baryon 
and of a hypothetical baryon state. 
This factor varies from $0.96\, (0.92)$ at the smallest masses to
$0.87\, (0.86)$ at the highest masses considered for the 
neutral (doubly charged) states.

The ratio $R(M)$  has the advantage that
the systematic effects of the acceptances and the reconstruction efficiencies
mostly cancel, making it insensitive to detector effects.
In the absence of an observed signal, the ratio $R(M)$ therefore provides 
a robust quantity to
set upper limits on the production of new narrow baryonic resonances
decaying to $\Xi^-\pi^{\pm}$ in the mass range $1600 - 2300 \mev$.
A mass-dependent upper limit at the $95\%$  confidence level (C.L.) on 
the ratio $R(M)$ is obtained from the observed invariant mass spectra
using a modified frequentist approach 
based on likelihood ratios \cite{tjunk} analogous 
to the one applied in \cite{Aktas:2006ic}. 
The method takes the 
statistical and systematic uncertainties in the 
number of signal events and background combinations into account.


The systematic uncertainties on the ratio  $R(M)$ 
include the following contributions:
\begin{enumerate}
\item The uncertainty on the overall number of \xizero candidates is 
$15\,\%$, as determined from the
fit to the mass spectra.
%
\item The statistical uncertainty on the determination of the 
reconstruction efficiency 
$\epsilon(M,q)$ by the Monte Carlo simulation is taken as the value 
of the systematic uncertainty on the ratio 
$\epsilon(1530,0)/\epsilon(M,q)$. 
It is found to be below $8\,\%$ for all masses and charge 
combinations considered. Systematic uncertainties 
due to the lack of knowledge of the production mechanism 
of the hypothetical new baryon state are not taken into 
account.
\item An uncertainty on the experimental 
width of the potential pentaquark signal is 
estimated to be $5\,\%$, as determined by the
difference of the observed width of the \xizero\ 
signal in data and simulation.
\item A systematic uncertainty on the background distribution is assessed
by performing the fit under different assumptions:
using the background function (\ref{bgeqn}) in the full mass range, 
excluding
a mass window  of $\pm 2 \sigma$ around the considered mass,
and also using the sum of the
background function and a Gaussian with fixed mass $M$
and width $\sigma(M)$ to account for a possible signal.
The uncertainty of the number of background combinations
is estimated from the difference between the different fitting methods and
amounts to $2\,\%$.

\end{enumerate}
In the limit calculation, the contributions from 
items 1-3 are quadratically added and included
in the uncertainty on the signal events, while the 
contribution from item 4 enters the background uncertainty.


The final results of the limit calculations
are quoted for the kinematic region $2 < Q^2 < 100\gev^2$ and $0.05 < y < 0.7$,
for $\pt(\Xi \pi) > 1\gev$  and $|\eta(\Xi \pi)|< 1.5$.
As stated before, it is assumed that new resonances are
produced by a similar mechanism to that of the known $J=3/2$ baryons, 
that they decay into $\Xi \pi$ with a $100\%$ branching ratio,
and that their natural widths are below the experimental resolution.

In figure~\ref{pentamm_limits}
the $\Xi^- \pi^-$ invariant mass spectrum summed over
the two doubly charged combinations is shown.
In the same figure the $95\%$ C.L. upper limit on the ratio  $R(M)$ is presented
as a function of the mass $M$.
The non-observation of a resonance state 
in the mass range $1600 - 2300 \mev$  
limits the production rate of a hypothetical 
\pqqc\ pentaquark to $12-45\%$ 
of the \xizero production rate
at the $95\%$ C.L., depending on the ($\Xi \pi$)-mass.

Furthermore, no signal is observed  in the neutral
invariant mass spectrum 
(figure~\ref{pentamp_limits}) in the mass range $1600 - 2300 \mev$, 
above the \xizero baryon. 
Figure \ref{pentamp_limits} also shows the resulting upper limit 
on the ratio $R(M)$, which
is obtained in a procedure analogous to the 
doubly charged case.
This restricts the production rate of a hypothetical  $\Xi^{0}_{5q}$ pentaquark state
to less than $10-50\%$ 
of that of the \xizero baryon,
depending on the ($\Xi \pi$)-mass.

This result is similar to the $10- 50\%$ upper limits measured
by the ZEUS Collaboration~\cite{Chekanov:2005at}
in the region $Q^2 > 1 \GeVSq$.

\section{Conclusion}

A search for new narrow baryonic resonances
decaying into $\Xi^{-}\pi^{-}$ 
and $\Xi^{-}\pi^{+}$ and their charge conjugate states 
is performed with the H1 detector using a DIS 
data sample in the kinematic region $2 < Q^{2} < 100\gev^{2}$ and $0.05 < y < 0.7$,
corresponding to a total  integrated luminosity of $101\,\pbinv$. 
The established \xizero baryon state is observed.
No signal of a new baryonic state is found in the mass range
$1600 - 2300 \mev$. Therefore
mass dependent upper limits at the $95\%$ C.L. are set on the production ratio of  
hypothetical states, such as \pqqc and \pqqn, to the  total number of
observed \xizero baryons.

The results reported here are similar to the limits measured
by the ZEUS Collaboration.
The overall H1 statistics in the $\Xi^-$ samples is comparable 
with the NA49 collaboration's data.
The H1 limits obtained at HERA do not confirm the NA49 
observation of potential pentaquark states.

\section*{Acknowledgements}

We are grateful to the HERA machine group whose outstanding
efforts have made this experiment possible. 
We thank the engineers and technicians for their work in constructing and
maintaining the H1 detector, our funding agencies for 
financial support, the
DESY technical staff for continual assistance
and the DESY directorate for support and for the
hospitality which they extend to the non DESY 
members of the collaboration.


\clearpage


\newpage

\clearpage
\newpage

\clearpage

\newpage

\newpage

\end{document}